# Size-dependent theories of piezoelectricity:
# Comparisons and further developments for centrosymmetric dielectrics

Ali R. Hadjesfandiari

*Department of Mechanical and Aerospace Engineering*
*University at Buffalo, State University of New York*
*Buffalo, NY 14260 USA*

ah@buffalo.edu

September 2, 2014

**Abstract**

Here the recently developed size-dependent piezoelectricity and the strain gradient theory of flexoelectricity are compared. In the course of this investigation, the strain gradient theory of flexoelectricity is shown to violate fundamental rules of mathematics, continuum mechanics and electromagnetism. The major difficulties are associated with ill-posed boundary conditions, the missing angular (moment) equilibrium equation and the appearance of a non-physical extraneous vectorial electrostatic law. Therefore, the strain gradient flexoelectricity must be classified as an inconsistent theory. The present investigation further reveals that the new size-dependent piezoelectricity is the more consistent theory to describe linear electromechanical coupling in dielectrics. Some new aspects of this theory are presented for isotropic and centrosymmetric cubic dielectric materials, whose coupling effect is described by only one parameter.

**1. Introduction**

What is considered today as the strain gradient theory of flexoelectricity is actually one of several existing theories in size-dependent piezoelectricity. This accepted flexoelectricity theory is based on the assumption that the electric polarization can be generated as the result of coupling to the third order strain gradient tensor $\frac{\partial e_{ij}}{\partial x_k}$ (Tagantsev, 1986, Maranganti et al., 2006; Eliseev et al., 2009). In a recent review paper (Yudin and Tagantsev, 2014), authors state that despite the considerable theoretical and experimental studies of flexoelectricity, there exist many open issues related to a limited understanding of the physics of flexoelectricity.

In this paper, we inspect strain gradient flexoelecticity from different angles, such as mathematical, mechanical and electrical. First we notice that in this theory the possible existence of couple-stresses $\mu_{ij}$ has not been explicitly considered. However, based on the rules of continuum mechanics, any second order gradient theory requires a second order tensor as a measure of deformation (Mindlin and Tiersten, 1962; Koiter, 1964). This in turn necessiates the appearance of couple-stresses $\mu_{ij}$ directly in the formulation. Therefore, a consistent size-dependent piezoelectricity theory must be compatible with this requirement and follow the consistent couple stress theory (Hadjesfandiari and Dargush, 2011). However, the present strain gradient flexoelectric theory does not satisfy this fundamental pillar of continuum mechanics. Therefore, the present state-of-the-art flexoelectricity theory must be questioned as a viable description of reality. Furthermore, by inspecting this theory more closely, some mathematical and physical inconsistencies are found, which will be explained here by comparing to the new development in size-dependent piezoelectricity (Hadjesfandiari, 2013). Interestingly, it will be shown that the present state-of-the-art flexoelectric theory also involves the creation of a new unsubstantiated electrostatic law. As a result, one is led to the conclusion that the new size-dependent piezoelectricity (Hadjesfandiari, 2013) is the consistent theory required to describe fully linear electromechanical coupling.

The paper is organized as follows. In Section 2, we provide an overview of the consistent size-dependent piezoelectric theory. This section also includes the formulation for centrosymmetric dielectric materials. In Section 3, we investigate the strain gradient flexoelectric theory and demonstrate its mathematical, mechanical and electromagnetical inconsistencies. This includes the disturbing appearance of a non-physical extraneous electrostatic law, which is the result of considering the contribution from the gradient of polarization in the energy density function. Next, we compare the size-dependent piezoelectric and the inconsistent flexoelectric theories, and investigate their interrelationships in Section 4. This illustrates the importance of the recent development in couple stress theory and demonstrates that the new size-dependent piezoelectricity is in fact the consistent flexoelectricity theory. Section 5 presents some important aspects of the consistent size-dependent piezoelectricity theory for isotropic materials. Section 6 presents a consistent couple stress flexoelectricity theory for centrosymmetric cubic dielectric materials. Finally, Section 7 contains a summary and some general conclusions.

## 2. Consistent size-dependent piezoelectric theory

The consistent size-dependent piezoelectricity theory (Hadjesfandiari, 2013) is based on the recent development in couple stress theory (Hadjesfandiari and Dargush, 2011), which establishes that the displacement $u_i$ and rotation $\omega_i$ are the kinematical degrees of freedom, since these are energetically conjugate to force traction $t_i^{(n)}$ and moment traction $m_i^{(n)}$, respectively. As a result, the internal stresses are represented by true force-stress $\sigma_{ij}$ and pseudo couple-stress $\mu_{ij}$ tensors. The triumph of this development is the discovery of the subtle skew-symmetric character of the couple-stress tensor

$$\mu_{ji} = -\mu_{ij} \tag{1}$$

The components of these force-stress and couple-stress tensors are shown in Fig. 1. The skew-symmetric character of the couple-stress tensor resolves the serious difficulties that existed within the original couple stress theory (Cosserat and Cosserat, 1909) for more than a century.

The consistent size-dependent continuum mechanics is a practical theory, which enables us to develop different size-dependent formulations in many multi-physics disciplines, such as electromechanics.

In this theory, the consistent curvature tensor is the mean curvature tensor, which is the skew-symmetrical part of the rotation gradient, that is

$$\kappa_{ij} = \omega_{[i,j]} = \frac{1}{2}\left(\omega_{i,j} - \omega_{j,i}\right) \tag{2}$$

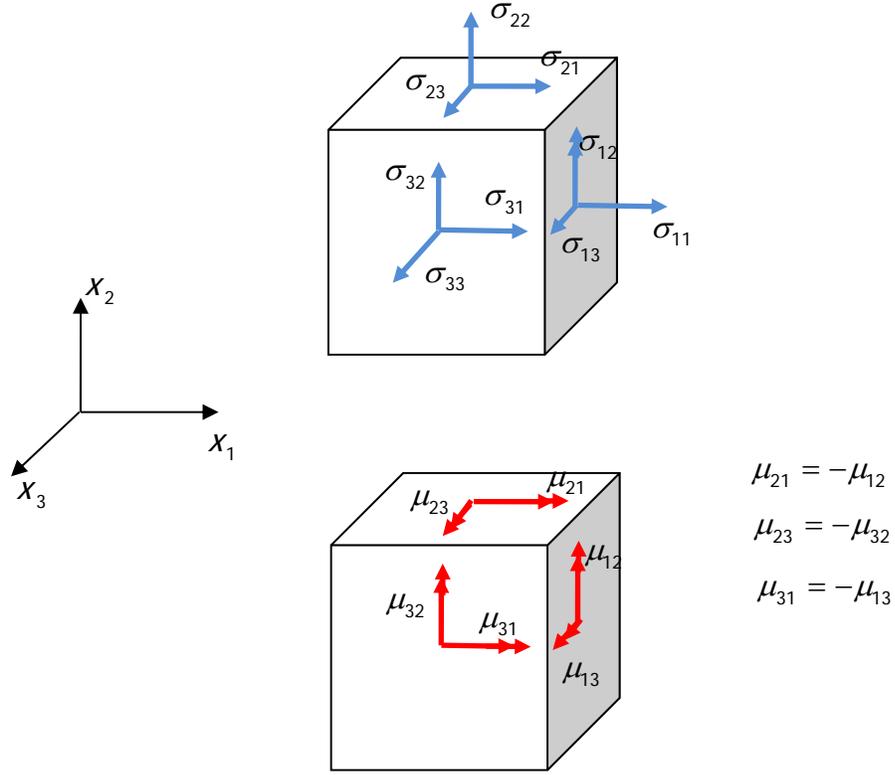

**Fig. 1.** Components of force- and couple-stress tensors in the present consistent theory.

It should be noticed that the strain and rotation tensors are defined as

$$e_{ij} = \frac{1}{2}\left(u_{i,j} + u_{j,i}\right) \tag{3}$$

$$\omega_{ij} = \frac{1}{2}\left(u_{i,j} - u_{j,i}\right) \tag{4}$$

respectively. The rotation vector $\omega_i$ is related to the rotation tensor as

$$\omega_i = \frac{1}{2}\varepsilon_{ijk}\omega_{kj} \tag{5}$$

The true couple-stress vector $\mu_i$ and true mean curvature vector $\kappa_i$ dual to their corresponding tensors are defined by

$$\mu_i = \frac{1}{2}\varepsilon_{ijk}\mu_{kj} \tag{6}$$

$$\kappa_i = \frac{1}{2}\varepsilon_{ijk}\kappa_{kj} \tag{7}$$

where we also have

$$\varepsilon_{ijk}\mu_k = \mu_{ji} \tag{8}$$

$$\varepsilon_{ijk}\kappa_k = \kappa_{ji} \tag{9}$$

The force-stress tensor $\sigma_{ji}$ is generally non-symmetric and can be decomposed as

$$\sigma_{ji} = \sigma_{(ji)} + \sigma_{[ji]} \tag{10}$$

where $\sigma_{(ji)}$ and $\sigma_{[ji]}$ are the symmetric and skew-symmetric parts, respectively. We can consider the axial vector $s_i$ dual to $\sigma_{[ij]}$, where

$$s_i = \frac{1}{2}\varepsilon_{ijk}\sigma_{[kj]} \tag{11}$$

which also satisfies

$$\varepsilon_{ijk}s_k = \sigma_{[ji]} \tag{12}$$

For a quasistatic electric field $E_i$, in which the effect of induced magnetic field in the material is neglected, we have the electrostatic relation

$$\varepsilon_{ijk}E_{k,j} = 0 \tag{13}$$

Therefore, the electric field $E_i$ can be represented by the electric potential $\phi$, such that

$$E_i = -\phi_{,i} \tag{14}$$

The electric field and deformation can induce polarization $P_i$ in the dielectric material. The electric displacement vector $D_i$ is defined by

$$D_i = \epsilon_0 E_i + P_i \tag{15}$$

where $\epsilon_0$ is the permittivity of free space.

The fundamental electromechanical equations for a static size-dependent piezoelectric theory (Hadjesfandiari, 2013) are the generalized equilibrium equations

$$\sigma_{ji,j} + F_i = 0 \qquad \text{Linear balance law} \qquad (16)$$

$$\mu_{ji,j} + \varepsilon_{ijk}\sigma_{jk} = 0 \qquad \text{Angular balance law} \qquad (17)$$

$$D_{i,i} = \rho_e \qquad \text{Gauss law} \qquad (18)$$

where $F_i$ is the body force per unit volume, and $\rho_e$ is the electric charge density in the volume of the body. The force-traction vector $t_i^{(n)}$, moment-traction vector $m_i^{(n)}$, and normal electric displacement $d$ at a point on surface element $dS$ with unit outward normal vector $n_i$ are given by

$$t_i^{(n)} = \sigma_{ji} n_j \qquad (19)$$

$$m_i^{(n)} = \mu_{ji} n_j = \varepsilon_{ijk} n_j \mu_k \qquad (20)$$

$$d = D_i n_i \qquad (21)$$

respectively. Since the couple-stress tensor is skew-symmetric, the moment-traction vector $m_i^{(n)}$ is tangent to the surface element. The force-traction $t_i^{(n)}$ and the consistent bending moment-traction $m_i^{(n)}$ acting on an arbitrary surface with unit normal vector $n_i$ are shown in Fig. 2.

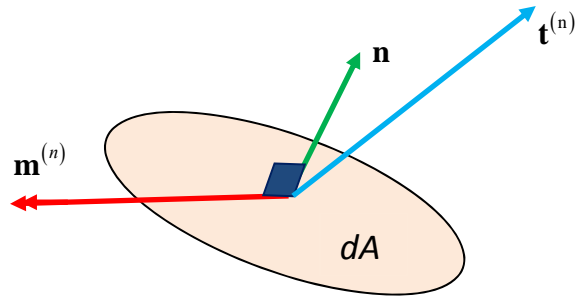

**Fig. 2.** Force-traction $\mathbf{t}^{(n)}$ and the consistent bending moment-traction $\mathbf{m}^{(n)}$.

The angular or moment equilibrium Eq. (17) gives the skew-symmetric part of the force-stress tensor as

$$\sigma_{[ji]} = -\mu_{[i,j]} = -\frac{1}{2}(\mu_{i,j} - \mu_{j,i}) \tag{22}$$

Therefore, for the total force-stress tensor, we have

$$\sigma_{ji} = \sigma_{(ji)} + \sigma_{[ji]} = \sigma_{(ji)} + \mu_{[j,i]} \tag{23}$$

As a result, the system of equations reduce to

$$[\sigma_{(ji)} + \mu_{[j,i]}]_{,j} + F_i = 0 \qquad \text{Linear balance law} \tag{24}$$

$$D_{i,i} = \rho_e \qquad \text{Gauss law} \tag{25}$$

It is seen that the sole duty of the angular equilibrium Eq. (17) is to produce the skew-symmetric part of the force-stress tensor.

The natural electromechanical boundary conditions are two mechanical and one electrical boundary condition

$$t_i^{(n)} = \bar{t}_i^{(n)} \qquad \text{on } S_t \tag{26}$$

$$m_i^{(n)} = \bar{m}_i^{(n)} \qquad \text{on } S_m \tag{27}$$

$$d = \bar{d} \qquad \text{on } S_d \tag{28}$$

where $S_t$, $S_m$ and $S_d$ are the portions of surface at which $t_i^{(n)}$, $m_i^{(n)}$ and $d$ are prescribed, respectively. The essential or geometrical boundary conditions are

$$u_i = \bar{u}_i \qquad \text{on } S_u \tag{29}$$

$$\omega_i = \bar{\omega}_i \qquad \text{on } S_\omega \tag{30}$$

$$\phi = \bar{\phi} \qquad \text{on } S_\phi \tag{31}$$

where $S_u$, $S_\omega$ and $S_\phi$ are the portions of surface at which $u_i$, $\omega_i$ and $\phi$ are prescribed, respectively. In Eqs. (26)-(31), the over bar is used to denote a specified quantity.

It should be noticed that since the moment-traction vector $m_i^{(n)}$ is tangent to the boundary surface, the normal component of $\omega_i$ cannot be considered as an independent degree of freedom. Therefore, only the tangential component of $\omega_i$ can be prescribed on $S_\omega$.

Hadjesfandiari (2013) has shown that the specific electric enthalpy $H$ for a size-dependent piezoelectric material depends on the strain tensor $e_{ij}$, the mean curvature vector $\kappa_{ij}$, and the electric field vector $E_i$, that is

$$H = H(e_{ij}, \kappa_i, E_i) \qquad (32)$$

As a result, we obtain the following constitutive relations

$$\sigma_{(ji)} = \frac{1}{2}\left(\frac{\partial H}{\partial e_{ij}} + \frac{\partial H}{\partial e_{ji}}\right) \qquad (33)$$

$$\mu_i = -\frac{1}{2}\frac{\partial H}{\partial \kappa_i} \qquad (34)$$

$$D_i = -\frac{\partial H}{\partial E_i} \qquad (35)$$

Furthermore, from Eq. (22) we derive

$$\sigma_{[ji]} = -\mu_{[i,j]} = \frac{1}{4}\left(\frac{\partial H}{\partial \kappa_i}\right)_{,j} - \frac{1}{4}\left(\frac{\partial H}{\partial \kappa_j}\right)_{,i} \qquad (36)$$

As a result, for the total force-stress tensor we have

$$\sigma_{ji} = \frac{1}{2}\left(\frac{\partial H}{\partial e_{ij}} + \frac{\partial H}{\partial e_{ji}}\right) + \frac{1}{4}\left[\left(\frac{\partial H}{\partial \kappa_i}\right)_{,j} - \left(\frac{\partial H}{\partial \kappa_j}\right)_{,i}\right] \qquad (37)$$

For linear elastic theory, we consider the homogeneous quadratic form for the enthalpy density function as

$$H = H\left(e_{ij}, \kappa_i, E_i\right) = \frac{1}{2} A_{ijkl} e_{ij} e_{kl} + \frac{1}{2} B_{ij} \kappa_i \kappa_j + C_{ijk} e_{ij} \kappa_k - \frac{1}{2} \epsilon_{ij} E_i E_j - \alpha_{ijk} E_i e_{jk} - \beta_{ij} E_i \kappa_j \quad (38)$$

where $A_{ijkl}$, $B_{ijkl}$ and $C_{ijkl}$ are the elasticity tensors; $\epsilon_{ij}$ is the permittivity tensor; and $\alpha_{ijk}$ and $\beta_{ij}$ are piezoelectricity and flexoelectricity tensors, respectively (Hadjesfandiari, 2013). The symmetry relations are

$$A_{ijkl} = A_{klij} = A_{jikl} \quad (39)$$

$$B_{ijkl} = B_{klij} = -B_{jikl} \quad (40)$$

$$C_{ijkl} = C_{jikl} = -C_{ijlk} \quad (41)$$

$$\epsilon_{ij} = \epsilon_{ji} \quad (42)$$

$$\alpha_{ijk} = \alpha_{ikj} \quad (43)$$

$$\beta_{ij} = \beta_{ji} \quad (44)$$

Therefore, by using Eqs. (33)-(36), we obtain

$$\sigma_{(ji)} = A_{ijkl} e_{kl} + C_{ijk} \kappa_k - \alpha_{kji} E_k \quad (45)$$

$$\mu_i = -\frac{1}{2} B_{ij} \kappa_j - \frac{1}{2} C_{kji} e_{kj} + \frac{1}{2} \beta_{ji} E_j \quad (46)$$

$$D_i = \epsilon_{ij} E_j + \alpha_{ijk} e_{jk} + \beta_{ij} \kappa_j \quad (47)$$

$$\sigma_{[ji]} = \frac{1}{4} B_{im} \kappa_{m,j} - \frac{1}{4} B_{jm} \kappa_{m,i} + \frac{1}{4} C_{kmi} e_{km,j} - \frac{1}{4} C_{kmj} e_{km,i} - \frac{1}{4} \beta_{mi} E_{m,j} + \frac{1}{4} \beta_{mj} E_{m,i} \quad (48)$$

for a homogeneous anisotropic dielectric material. It should be noticed that although there is a coupling between the gradient of the electric field $E_{i,j}$ and the skew-symmetric part of the force-stress $\sigma_{[ji]}$ tensor, there is no term associated with $E_{i,j}$ in the enthalpy density $H$ and the constitutive relations for $\sigma_{(ji)}$, $\mu_i$ and $D_i$. The appearance of $E_{i,j}$ in the skew-symmetric part of force-stress tensor $\sigma_{[ji]}$ is because of the angular balance law Eq. (17), which is not a constitutive equation.

By carrying the constitutive relations into the linear equilibrium Eq. (24) and electric Gauss law Eq. (25), we obtain

$$A_{ijkl}u_{k,lj} + \frac{1}{4}C_{ijk}\left(u_{m,mjk} - \nabla^2 u_{k,j}\right) + \frac{1}{4}C_{kmi}\nabla^2 u_{k,m} - \frac{1}{4}C_{kmj}u_{k,mij}$$
$$+ \frac{1}{16}B_{ik}\left(\nabla^2 u_{m,mk} - \nabla^2\nabla^2 u_k\right) - \frac{1}{16}B_{jk}\left(u_{m,mkij} - \nabla^2 u_{k,ij}\right) \quad (49)$$
$$+ \alpha_{kji}\phi_{,kj} + \frac{1}{4}\beta_{mi}\nabla^2\phi_{,m} - \frac{1}{4}\beta_{mj}\phi_{,mij} + F_i = 0$$

$$\epsilon_{ij}\phi_{,ij} - \alpha_{ijk}u_{j,ik} - \frac{1}{4}\beta_{ij}\left(u_{m,mij} - \nabla^2 u_{j,i}\right) + \rho_e = 0 \quad (50)$$

In terms of displacements, the linear balance Eq. (49) is a fourth order vectorial equation in space, which requires specification of two vectorial boundary conditions. Therefore, we can specify either the displacement vector $u_i$ or the force-traction vector $t_i^{(n)}$, the tangential component of the rotation vector $\omega_i$ or the tangent moment-traction vector $m_i^{(n)}$, and the electric potential $\phi$ or the normal electric displacement $d$. These six boundary conditions are given in Eqs. (26)-(31).

The fundamental governing equations and the boundary conditions for size-dependent piezoelectricity or consistent flexoelectric theory can also be obtained by different variational methods. This is demonstrated in Appendix A by extremizing the total energy functional based on the enthalpy density function.

Consequently, by considering different aspects of newly developed size-dependent piezoelectricity, we conclude that this theory is a consistent theory of flexoelectricity, which has the potential to represent physical reality.

## 2.1. Centrosymmetric dielectric materials

For centrosymmetric dielectric materials, we have

$$C_{ijk} = 0 \quad (51)$$

$$\alpha_{ijk} = 0 \quad (52)$$

This shows that there is no classical piezoelectricity in the material and the enthalpy density function reduces to

$$H = H\left(e_{ij}, \kappa_i, E_i\right) = \frac{1}{2} A_{ijkl} e_{ij} e_{kl} + \frac{1}{2} B_{ij} \kappa_i \kappa_j - \frac{1}{2} \epsilon_{ij} E_i E_j - \beta_{ij} E_i \kappa_j \tag{53}$$

Therefore, the constitutive relations become

$$\sigma_{(ji)} = A_{ijkl} e_{kl} \tag{54}$$

$$\mu_i = -\frac{1}{2} B_{ij} \kappa_j + \frac{1}{2} \beta_{ji} E_j \tag{55}$$

$$D_i = \epsilon_{ij} E_j + \beta_{ij} \kappa_j \tag{56}$$

In addition, for the skew-symmetric part of the force-stress tensor, we obtain

$$\sigma_{[ji]} = \frac{1}{4} B_{im} \kappa_{m,j} - \frac{1}{4} B_{jm} \kappa_{m,i} - \frac{1}{4} \beta_{mi} E_{m,j} + \frac{1}{4} \beta_{mj} E_{m,i} \tag{57}$$

As a result, the governing equations (49) and (50) reduce to

$$A_{ijkl} u_{k,lj} + \frac{1}{16} B_{ik} \left(\nabla^2 u_{m,mk} - \nabla^2 \nabla^2 u_k\right) - \frac{1}{16} B_{jk} \left(u_{m,mkij} - \nabla^2 u_{k,ij}\right)$$
$$+ \frac{1}{4} \beta_{mi} \nabla^2 \phi_{,m} - \frac{1}{4} \beta_{mj} \phi_{,mij} + F_i = 0 \tag{58}$$

$$\epsilon_{ij} \phi_{,ij} - \frac{1}{4} \beta_{ij} \left(u_{m,mij} - \nabla^2 u_{j,i}\right) + \rho_e = 0 \tag{59}$$

## 3. Strain gradient flexoelectric theory and its inconsistencies

Now we examine the state-of-the-art theory of strain gradient flexoelectricity, which has been generally accepted as the correct theory by mainstream investigators. In this formulation, which does not take account of the physical couple-stresses $\mu_{ij}$ directly, some fundamental mathematical and physical rules are violated. Furthermore, this theory, which is examined based on the work of Maranganti et al. (2006), includes the appearance of a new electrostatic law along with the Gauss law.

In this flexoelectric theory, the internal energy density function is assumed to be in the form of

$$\Sigma = \Sigma\left(e_{ij}, u_{j,kl}, P_i, P_{i,j}\right) \tag{60}$$

with the most general expression

$$\Sigma = \frac{1}{2}a_{ij}P_iP_j + \frac{1}{2}b_{ijkl}P_{i,j}P_{k,l} + \frac{1}{2}c_{ijkl}e_{ij}e_{kl} + d_{ijkl}P_{i,j}e_{kl} + f_{ijkl}P_i u_{j,kl} + \frac{1}{2}g_{ijklmn}u_{i,jk}u_{l,mn} \tag{61}$$

for linear theory. The appearance of the gradient of polarization $P_{i,j}$ cannot be justified physically or mathematically and creates strange consequences, as we shall see later.

Based on the energy density function Eq. (61), the fundamental equations have been given as

$$\left(t_{ij} - t_{jim,m}\right)_{,j} + F_i = 0 \quad \text{Linear balance law} \tag{62}$$

$$E_{ij,j} + E_i - \phi_{,i} + E_i^0 = 0 \quad \text{Unknown law} \tag{63}$$

$$-\epsilon_0 \phi_{,ii} + P_{i,i} = 0 \quad \text{Gauss law} \tag{64}$$

where the tensor $t_{ij}$ is the stress tensor in classical elasticity. The terms $t_{jim,m}$ and $E_{ij}$ have been thought of as higher order stress (moment stress) and higher order local electric force, respectively. (These notations have been used in Maranganti et al., 2006.) As a result, the term $t_{ij} - t_{jim,m}$ has been interpreted as a non-symmetric physical stress

$$\sigma_{ij}^{phys} = t_{ij} - t_{jim,m} \tag{65}$$

We notice that Eqs. (62) and (64) intend to represent the linear momentum balance and Gauss law, respectively, for quasistatic response. Although Eq. (63) cannot be recognized as a Maxwell's equation, it has an electrical character. As a result, there must be also one mechanical and two electrical natural boundary conditions

$$n_i \sigma_{ij}^{phys} = t_j \tag{66}$$

$$E_{ij} n_i = 0 \tag{67}$$

$$n_i \left(\lfloor \epsilon_0 \phi_{,i} \rfloor + P_i \right) = 0 \tag{68}$$

where the symbol $\lfloor \ \rfloor$ denotes the jump across the surface.

We notice that there is no angular balance equation and its corresponding boundary conditions $\omega_i$ or $m_i^{(n)}$. In terms of displacements, the linear balance Eq. (62) is fourth order, which then requires two vectorial boundary conditions to ensure a compatible well-posed boundary value problem. However, in this flexoelectricity theory, we can only specify either the displacement vector $u_i$ or the force-traction vector $n_i \sigma_{ij}^{phys} = t_j$ Eq. (66) on the boundary. There is no additional mechanical boundary condition, but instead two electrical boundary conditions Eqs. (67) and (68) appear. The apparently new electrical balance law Eq. (63) and its corresponding natural boundary condition $E_{ij} n_i = 0$ is bewildering. Furthermore, a careful examination shows that this new balance law Eq. (63) cannot be matched with any of the existing Maxwell's equations. The appearance of one mechanical law Eq. (62) and two electrical balance laws Eqs. (63) and (64) indicates that the formulation suffers from serious inconsistencies. Therefore, the strain gradient theory of flexoelectricity violates fundamental aspects of mathematics, continuum mechanics and also electromagnetism.

These violations can be elucidated by decoupling the mechanical and electrical parts as follows. By neglecting the coupling between polarization and deformation, i.e., $d_{ijkl} = 0$ and $f_{ijkl} = 0$, we must obtain separate mathematically consistent elasticity and physically acceptable electrostatic polarization formulations. Based on what has been explained above, we notice that the resulting strain gradient elasticity does not even satisfy basic mathematical requirements. While the governing equation is the fourth order linear Eq. (62), there is only one mechanical natural boundary condition (66). This means that in this strain gradient elasticity the specification of boundary conditions is similar to those in the classical elasticity, in which one mechanical boundary condition $t_j$ or one kinematical boundary condition $u_j$ are prescribed. However, the fourth order of the governing equation requires a second set of boundary conditions, which does not exist in this formulation. Furthermore, the resulting electrical polarization is governed by two Eqs. (63) and (64) with boundary conditions prescribed by Eqs. (67) and (68). However, we expect the polarization must follow the classical electrostatic theory and be governed only by Eq.

(64) with natural boundary condition Eq. (68). This means the gradient of polarization $P_{i,j}$ should not appear in the formulation.

It should be noticed that the inconsistent new electrical balance law Eq. (63) is actually due to Mindlin (1968), who considered the appearance of the gradient of the electric field or polarization in the enthalpy density function. Interestingly, we should notice that he did not include any form of second gradient of deformation in his formulation. This clearly shows that he was not certain about the validity of any of the second gradient theories that he was developing at the time. These include indeterminate couple stress theory (Mindlin and Tiersten, 1962), strain gradient theory (Mindlin and Eshel, 1968), and micropolar theory (Mindlin, 1964). Nevertheless, it is interesting to note that at the present not only the troubles with the couple stress theory has been resolved (Hadjesfandiari and Dargush, 2011), the consistent flexoelectricity theory has also been developed (Hadjesfandiari, 2013).

## 4. Interrelationship of the size-dependent piezoelectric and strain gradient flexoelectric theories

Interestingly, the inconsistent mathematical formulation of the strain gradient flexoelectricity possesses some hidden clues, which could have given some hints to its proponents. By comparing the equations in this theory with the current consistent theory, we notice that the former flexoelectricity formulation resembles the consistent couple stress theory, with the corresponding analogies

$$\sigma_{ij}^{phys} \leftrightarrow \sigma_{ji} \tag{69}$$

$$t_{ij} \leftrightarrow \sigma_{(ji)} \tag{70}$$

$$-t_{jim,m} \leftrightarrow \sigma_{[ji]} = -\mu_{[i,j]} \tag{71}$$

This means that in the former strain gradient flexoelectricty, the tensors $t_{ij}$ and $-t_{jim,m}$ would resemble the roles of symmetric and skew-symmetric parts of the force stress tensors $\sigma_{(ji)}$ and $\sigma_{[ji]}$, respectively. However, $-t_{jim,m}$ is not skew-symmetric and has been taken ambiguously as a higher order stress or moment stress (Maranganti et al., 2006).

It is seen that by considering the third order strain gradient tensor $\dfrac{\partial e_{ij}}{\partial x_k}$ as the measure of deformation and neglecting the existence of couple stresses, the role of $-t_{jim,m}$ as the skew-symmetric part of the force-stress tensor has not been recognized. Furthermore, the requirement to prescribe the necessary mechanical boundary conditions $\omega_i$ or $m_i^{(n)}$ has been missed. Nevertheless, this missing mechanical boundary condition has been compensated by an apparent electrical based boundary condition. This is because the theory involves two electrical laws

$$E_{ij,j} + E_i - \phi_{,i} + E_i^0 = 0 \qquad \text{Unknown law} \qquad (72)$$

$$-\epsilon_0 \phi_{,ii} + P_{i,i} = 0 \qquad \text{Gauss law} \qquad (73)$$

The second equation is recognized to be the Gauss law, but the first vectorial equation has no corresponding fundamental law in Maxwell's equations. This clearly shows that Eq. (72) is completely unsubstantiated. There should not be a direct coupling to the gradient of polarization $P_{i,j}$ in the internal energy density function $\Sigma$. Now it can be realized that at most, we have $\Sigma = \Sigma\left(e_{ij}, u_{j,kl}, P_i\right)$. We also notice that the quantities $d = D_i n_i$ and $\phi$ are energy conjugate, but there is no physical energy conjugate for $E_{ij} n_i$. Furthermore, the strain gradient flexoelectricity does not recognize the existence of the physical angular balance law, but instead creates a new electrical balance law by violating Maxwell's equations. It is not surprising to see that researchers are still struggling with the issue of boundary conditions (Yurkov, 2011), which demonstrates the inconsistency of the theory of strain gradient flexoelectricity from the beginning. As demonstrated above, this theory reduces to a strain gradient elasticity with a missing set of boundary conditions.

It should be mentioned that there have been other piezoelectric theories (i.e., Wang et al., 2004), which consider the possible existence of physical couple stresses. However, these developments suffer from their dependence on underlying inconsistent couple stress theories. It can be said that these theories are somewhat more consistent, because at least they consider the existence of couple-stresses.

## 5. Isotropic linear couple stress flexoelectric dielectric materials

The flexoelectric effect in isotropic dielectric materials has already been studied by Hadjesfandiari (2013), where the constitutive tensors are

$$A_{ijkl} = \Lambda \delta_{ij}\delta_{kl} + G\delta_{ik}\delta_{jl} + G\delta_{il}\delta_{jk} \tag{74}$$

$$B_{ij} = 16\eta \delta_{ij} \tag{75}$$

$$C_{ijk} = 0 \tag{76}$$

$$\epsilon_{ij} = \epsilon \delta_{ij} \tag{77}$$

$$\alpha_{ijk} = 0 \tag{78}$$

$$\beta_{ij} = 4f\delta_{ij} \tag{79}$$

Here the moduli $\Lambda$ and $G$ (shear modulus) have the same meaning as the Lamé constants for an isotropic material in Cauchy elasticity. These constants are related to Young modulus $E$ and Poisson ratio $\nu$ by the relations

$$\Lambda = \frac{\nu E}{(1-2\nu)(1+\nu)} \tag{80}$$

$$G = \frac{E}{2(1+\nu)} \tag{81}$$

The parameter $\epsilon$ is the permittivity of the dielectric, and the parameters $\eta$ and $f$ account for its couple stress and flexoelectricity effects. The ratio

$$\frac{\eta}{G} = l^2 \tag{82}$$

specifies the characteristic material length scale $l$, which accounts for size-dependency.

As a result, the constitutive relations become

$$\sigma_{ji} = \Lambda e_{kk}\delta_{ij} + 2Ge_{ij} + 2Gl^2\nabla^2\omega_{ji} \tag{83}$$

$$\begin{aligned}\mu_i &= -8\eta\kappa_i + 2fE_i \\ &= -8Gl^2\kappa_i + 2fE_i\end{aligned} \tag{84}$$

$$D_i = \epsilon E_i + 4f\kappa_i \tag{85}$$

It is interesting to note that there is no direct coupling between the electric field $E_i$ and the total force-stress tensor $\sigma_{ji}$, a character common with the classical isotropic case. The coupling is only between the electric field $E_i$ and the couple-stress vector $\mu_i$.

In addition, note that the flexoelectric effect in isotropic dielectric is specified only by one coefficient $f$. This is in contrast with the results from inconsistent strain gradient flexoelectric theories, which predict three flexoelectric coefficients. Therefore, experimentalists should look for one flexoelectric coefficient $f$ for isotropic dielectric material.

For the governing equations, we obtain

$$\left[\Lambda + G(1+l^2\nabla^2)\right]u_{k,ki} + G(1-l^2\nabla^2)\nabla^2 u_i + F_i = 0 \tag{86}$$

$$\epsilon\nabla^2\phi + \rho_e = 0 \tag{87}$$

where the normal tractions are

$$t_i^{(n)} = \sigma_{ji}n_j = \left(\Lambda e_{kk}\delta_{ij} + 2Ge_{ij} + 2Gl^2\nabla^2\omega_{ji}\right)n_j \quad \text{on } S \tag{88}$$

$$m_i^{(n)} = \varepsilon_{ijk}n_j\mu_k = \varepsilon_{ijk}n_j\left(-8Gl^2\kappa_k + 2fE_k\right) \quad \text{on } S \tag{89}$$

Notice that the governing equations are explicitly independent of $f$. However, the flexoelectric effect can exist due to the moment-traction Eq. (89), which couples $\kappa_i$ and $E_i$ on the boundary.

This means the deformation $u_i$ and electric potential $\phi$ are coupled through boundary condition Eq. (89).

## 6. Linear centrosymmetric cubic couple stress flexoelectric dielectric materials

Here we explore the effect of size-dependent piezoelectricity or consistent couple stress flexoelectricity in centrosymmetric cubic materials, which is important in experimental investigations. For a centrosymmetric cubic material, the constitutive tensors $B_{ij}$, $\epsilon_{ij}$ and $\beta_{ij}$ are the same as their corresponding tensors for isotropic materials. Therefore, we have

$$B_{ij} = 16\eta \delta_{ij} \tag{90}$$

$$C_{ijk} = 0 \tag{91}$$

$$\epsilon_{ij} = \epsilon \delta_{ij} \tag{92}$$

$$\alpha_{ijk} = 0 \tag{93}$$

$$\beta_{ij} = 4f\delta_{ij} \tag{94}$$

where the material parameters $\epsilon$, $\eta$ and $f$ have the same meaning as their counterparts for isotropic dielectric materials. It is important to notice that the flexoelectricity character of centrosymmetric cubic materials is isotropic and is specified by only one flexoelectric coefficient $f$. This is in contrast with the results from strain gradient flexoelectric theories, which predict three flexoelectric coefficients for cubic materials (Tagantsev, 1986; Maranganti et al., 2006; Eliseev et al., 2009).

For centrosymmetric cubic materials the elasticity tensor $A_{ijkl}$ is specified by three material constants $A_{11}$, $A_{12}$ and $A_{44}$. By choosing the coordinate axes $x_1 x_2 x_3$ perpendicular to the symmetry planes of cubic materials, the symmetry conditions lead to the non-zero components

$$A_{1111} = A_{2222} = A_{3333} = A_{11} \tag{95}$$

$$A_{1122} = A_{2233} = A_{3311} = A_{12} \tag{96}$$

$$A_{1212} = A_{2323} = A_{1313} = A_{44} \tag{97}$$

The constitutive relation for the symmetric part of force-stress tensor $\sigma_{(ij)}$ can be written as

$$\begin{bmatrix} \sigma_{11} \\ \sigma_{22} \\ \sigma_{33} \\ \sigma_{(23)} \\ \sigma_{(13)} \\ \sigma_{(12)} \end{bmatrix} = \begin{bmatrix} A_{11} & A_{12} & A_{12} & 0 & 0 & 0 \\ A_{12} & A_{11} & A_{12} & 0 & 0 & 0 \\ A_{12} & A_{12} & A_{11} & 0 & 0 & 0 \\ 0 & 0 & 0 & A_{44} & 0 & 0 \\ 0 & 0 & 0 & 0 & A_{44} & 0 \\ 0 & 0 & 0 & 0 & 0 & A_{44} \end{bmatrix} \begin{bmatrix} e_{11} \\ e_{22} \\ e_{33} \\ 2e_{23} \\ 2e_{13} \\ 2e_{12} \end{bmatrix} \tag{98}$$

Accordingly, the inverse of this relation can be written as

$$\begin{bmatrix} e_{11} \\ e_{22} \\ e_{33} \\ 2e_{23} \\ 2e_{13} \\ 2e_{12} \end{bmatrix} = \begin{bmatrix} 1/E & -\nu/E & -\nu/E & 0 & 0 & 0 \\ -\nu/E & 1/E & -\nu/E & 0 & 0 & 0 \\ -\nu/E & -\nu/E & 1/E & 0 & 0 & 0 \\ 0 & 0 & 0 & 1/G & 0 & 0 \\ 0 & 0 & 0 & 0 & 1/G & 0 \\ 0 & 0 & 0 & 0 & 0 & 1/G \end{bmatrix} \begin{bmatrix} \sigma_{11} \\ \sigma_{22} \\ \sigma_{33} \\ \sigma_{(23)} \\ \sigma_{(13)} \\ \sigma_{(12)} \end{bmatrix} \tag{99}$$

This relation is virtually identical to the constitutive law for an isotropic solid, except that the shear modulus $G$ is not related to the Poisson ratio $\nu$ and Young modulus $E$ through the usual relation Eq. (81). The relationships among the elastic constants for cubic materials are

$$E = \left(A_{11}^2 + A_{12}A_{11} - 2A_{12}^2\right)/\left(A_{11} + A_{12}\right) \tag{100}$$

$$\nu = A_{12}/\left(A_{11} + A_{12}\right) \tag{101}$$

$$G = A_{44} \tag{102}$$

We can conveniently assume

$$\Lambda = A_{12} \tag{103}$$

As a result, we have

$$A_{11} = \frac{(1-\nu)E}{(1+\nu)(1-2\nu)} \tag{104}$$

$$A_{12} = \Lambda = \frac{\nu E}{(1+\nu)(1-2\nu)} \tag{105}$$

It is seen that in the special coordinate system $x_1 x_2 x_3$, the elasticity tensor can be represented as

$$A_{ijkl} = A_{12}\delta_{ij}\delta_{kl} + A_{44}(\delta_{ik}\delta_{jl} + \delta_{il}\delta_{jk}) + A_0 \delta_{ijkl} \qquad (106)$$

or

$$A_{ijkl} = \Lambda \delta_{ij}\delta_{kl} + G(\delta_{ik}\delta_{jl} + \delta_{il}\delta_{jk}) + A_0 \delta_{ijkl} \qquad (107)$$

where the tensor $\delta_{ijkl}$ is unity if all indices are alike and zero otherwise, and the coefficient $A_0$ is given by

$$A_0 = A_{11} - A_{12} - 2A_{44} = A_{11} - (\Lambda + 2G) \qquad (108)$$

The general transformation equations for the elasticity tensor from the orthogonal coordinate system $x_1 x_2 x_3$ to the new orthogonal coordinate system $x_1' x_2' x_3'$ is

$$A'_{ijkl} = a_{im} a_{jn} a_{kp} a_{lq} A_{mnpq} \qquad (109)$$

where $a_{ij}$ represents the orthogonal transformation matrix from the original system to the new system. Thomas (1966) has given the transformation of the elasticity tensor $A_{ijkl}$ as

$$A'_{ijkl} = A_{12}\delta_{ij}\delta_{kl} + A_{44}(\delta_{ik}\delta_{jl} + \delta_{il}\delta_{jk}) + A_0 a_{ib} a_{jb} a_{kb} a_{lb} \qquad (110)$$

where, we define the index $b = 1,2,3$ as a summation index.

Here we have demonstrated that the piezoelectric effect in a centrosymmetric cubic material is specified by only one flexoelectric coefficient $f$.

## 7. Conclusion

The generally accepted strain gradient theory of flexoelectricity is shown to violate not only the fundamental rules of continuum mechanics, but also the fundamental laws of electromagnetism. This creates an inconsistent flexoelectric boundary value problem, which misses the mechanical angular balance law, while introducing a new unsubstantiated electrostatic law. On the other hand, the recently developed size-dependent formulation, based on the consistent couple stress theory, gives a physically consistent flexoelectric theory. This new development has been considered for isotropic and centrosymmetric cubic materials, and some interesting results have been obtained.

**References**

Cosserat, E., Cosserat, F., 1909. Théorie des corps déformables (Theory of deformable bodies). A. Hermann et Fils, Paris.

Darrall, B. T., Dargush, G. F., Hadjesfandiari, A. R., 2014. Finite element Lagrange multiplier formulation for size-dependent skew-symmetric couple stress planar elasticity. Acta Mechanica, 225, 195-212.

Eliseev, E. A., Morozovska, A.N., Glinchuk, M. D., Blinc, R., 2009. Spontaneous flexoelectric/flexomagnetic effect in nanoferroics. Phys. Rev. B. 79, 165433.

Hadjesfandiari, A. R., 2013. Size-dependent piezoelectricity. Int. J. Solids Struct. 50, 2781-2791.

Hadjesfandiari, A. R., Dargush, G. F., 2011. Couple stress theory for solids. Int. J. Solids Struct. 48, 2496-2510.

Koiter, W. T., 1964. Couple stresses in the theory of elasticity, I and II. Proc. Ned. Akad. Wet. (B) 67, 17-44.

Maranganti, R., Sharma N. D., Sharma, P., 2006. Electromechanical coupling in nonpiezoelectric materials due to nanoscale nonlocal size effects: Green's function solutions and embedded inclusions. Phys. Rev. B. 74, 14110.

Mindlin, R. D., 1968. Polarization gradient in elastic dielectrics. Int. J. Solids Struct. 4, 637-642.

Mindlin, R.D., Eshel, N.N., 1968. On first strain-gradient theories in linear elasticity. Int. J. Solids Struct. 4, 109–124.

Mindlin, R.D., 1964. Micro-structure in linear elasticity. Arch. Ration. Mech. Anal.16, 51–78.

Mindlin, R. D., Tiersten, H. F., 1962. Effects of couple-stresses in linear elasticity, Arch. Rational Mech. Anal. 11, 415–488.

Tagantsev, A. K., 1986. Piezoelectricity and flexoelectricity in crystalline dielectrics. Phys. Rev. B. 34, 5883-5889.

Thomas, T.Y., 1966. On the stress-strain relations for cubic crystals. Proc. Natl. Acad. Sci. 55 (2), 235–239.

Wang, G.-F., Yu, S.-W., Feng, X.-Q., 2004. A piezoelectric constitutive theory with rotation gradient effects. Eur. J. Mech. A-Solid. 23, 455–466.

Yudin, P. V., Tagantsev, A. K., 2013. Fundamentals of flexoelectricity in solids. Nanotechnology. 24, 432001.

Yurkov, A. S., 2011. Elastic boundary conditions in the presence of the flexoelectric effect. Pis'ma v Zh. Èksper. Teoret. Fiz., 94, 490–493.

**Appendix A. Variational methods for the size-dependent piezoelectric theory**

Consider the general enthalpy density function as

$$H = H(e_{ij}, \kappa_i, E_i) \tag{A.1}$$

The total energy based on this enthalpy density for the electromechanical system is

$$\Pi_H = \int_V (H - F_i u_i + \rho_e \phi) dV - \int_{S_t} \overline{t}_i^{(n)} u_i dS - \int_{S_m} \overline{m}_i^{(n)} \omega_i dS - \int_{S_d} \overline{d} \phi dS \tag{A.2}$$

where $S_t$, $S_m$ and $S_d$ are the portions of the surface on which $t_i^{(n)}$, $m_i^{(n)}$ and $\phi$ are prescribed, respectively. Hadjesfandiari (2013) has already shown that the equilibrium condition corresponds to

$$\delta \Pi_H = 0 \tag{A.3}$$

where $\delta \Pi_H$ is the first variation of the functional $\Pi_H$. Therefore, we have

$$\delta \Pi_H = \int_V (\delta H - F_i \delta u_i + \rho_e \delta \phi) dV - \int_{S_t} \overline{t}_i^{(n)} \delta u_i dS - \int_{S_m} \overline{m}_i^{(n)} \delta \omega_i dS - \int_{S_d} \overline{d} \delta \phi dS \tag{A.4}$$

Note that $\bar{t}_i^{(n)}$, $\bar{m}_i^{(n)}$, $\bar{d}$, $F_i$ and $\rho_e$ are specified quantities, not subject to variation. By considering the conditions $\delta u_i = 0$ on $S_u$, $\delta \omega_i = 0$ on $S_\omega$ and $\delta \phi = 0$ on $S_\phi$, Eq. (A.4) can be written as

$$\delta \Pi_H = \int_V (\delta H - F_i \delta u_i + \rho_e \delta \phi) dV - \int_S t_i^{(n)} \delta u_i dS - \int_S m_i^{(n)} \delta \omega_i dS - \int_S d\delta \phi dS \tag{A.5}$$

As a result this becomes

$$\delta \Pi_H = \int_V \left( \frac{\partial H}{\partial e_{ij}} \delta e_{ij} + \frac{\partial H}{\partial \kappa_i} \delta \kappa_i + \frac{\partial H}{\partial E_i} \delta E_i - F_i \delta u_i + \rho_e \delta \phi \right) dV$$
$$- \int_S t_i^{(n)} \delta u_i dS - \int_S m_i^{(n)} \delta \omega_i dS - \int_S d\delta \phi dS \tag{A.6}$$

By some manipulation, we obtain

$$\delta \Pi_H = \int_V \left[ \frac{1}{2} \frac{\partial H}{\partial e_{ij}} (\delta u_{i,j} + \delta u_{j,i}) + \frac{1}{2} \frac{\partial H}{\partial \kappa_i} \delta \omega_{ji,j} - \frac{\partial H}{\partial E_i} \delta \phi_{,i} - F_i \delta u_i + \rho_e \delta \phi \right] dV$$
$$- \int_S t_i^{(n)} \delta u_i dS - \int_S m_i^{(n)} \delta \omega_i dS - \int_S d\delta \phi dS \tag{A.7}$$

or

$$\delta \Pi_H = \int_V \left[ \frac{1}{2} \left( \frac{\partial H}{\partial e_{ij}} + \frac{\partial H}{\partial e_{ji}} \right) \delta u_{i,j} + \frac{1}{2} \frac{\partial H}{\partial \kappa_i} \delta \omega_{ji,j} - \left( \frac{\partial H}{\partial E_i} \delta \phi \right)_{,i} + \left( \frac{\partial H}{\partial E_i} \right)_{,i} \delta \phi - F_i \delta u_i + \rho_e \delta \phi \right] dV$$
$$- \int_S t_i^{(n)} \delta u_i dS - \int_S m_i^{(n)} \delta \omega_i dS - \int_S d\delta \phi dS \tag{A.8}$$

This can also be written as

$$\delta \Pi_H = \int_V \left\{ \begin{array}{l} \left[ \frac{1}{2} \left( \frac{\partial H}{\partial e_{ij}} + \frac{\partial H}{\partial e_{ji}} \right) \delta u_i \right]_{,j} - \frac{1}{2} \left( \frac{\partial H}{\partial e_{ij}} + \frac{\partial H}{\partial e_{ji}} \right)_{,j} \delta u_i + \left[ \frac{1}{2} \frac{\partial H}{\partial \kappa_i} \delta \omega_{ji} \right]_{,j} - \frac{1}{2} \left( \frac{\partial H}{\partial \kappa_i} \right)_{,j} \delta \omega_{ji} \\ - \left( \frac{\partial H}{\partial E_i} \delta \phi \right)_{,i} + \left( \frac{\partial H}{\partial E_i} \right)_{,i} \delta \phi - F_i \delta u_i + \rho_e \delta \phi \end{array} \right\} dV$$
$$- \int_S t_i^{(n)} \delta u_i dS - \int_S m_i^{(n)} \delta \omega_i dS - \int_S d\delta \phi dS \tag{A.9}$$

or

$$\delta \Pi_H =$$

$$\int_V \left\{ \begin{array}{l} \left[ \dfrac{1}{2}\left( \dfrac{\partial H}{\partial e_{ij}} + \dfrac{\partial H}{\partial e_{ji}} \right) \delta u_i \right]_{,j} - \dfrac{1}{2}\left( \dfrac{\partial H}{\partial e_{ij}} + \dfrac{\partial H}{\partial e_{ji}} \right)_{,j} \delta u_i + \left[ \dfrac{1}{2}\dfrac{\partial H}{\partial \kappa_k} \varepsilon_{kji} \delta \omega_i \right]_{,j} - \dfrac{1}{2}\left( \dfrac{\partial H}{\partial \kappa_i} \right)_{,j} \delta \omega_{ji} \\ -\left( \dfrac{\partial H}{\partial E_i} \delta \phi \right)_{,i} + \left( \dfrac{\partial H}{\partial E_i} \right)_{,i} \delta \phi_i - F_i \delta u_i + \rho_e \delta \phi \end{array} \right\} dV \quad \text{(A.10)}$$

$$- \int_S t_i^{(n)} \delta u_i \, dS - \int_S m_i^{(n)} \delta \omega_i \, dS - \int_S d \delta \phi \, dS$$

Now we apply the divergence theorem to the first, third and fifth terms in the volume integral in Eq. (A.10) and obtain the relation

$$\delta \Pi_H = \int_V \left\{ \left[ -\dfrac{1}{2}\left( \dfrac{\partial H}{\partial e_{ij}} + \dfrac{\partial H}{\partial e_{ji}} \right)_{,j} + F_i \right] \delta u_i - \dfrac{1}{2}\left( \dfrac{\partial H}{\partial \kappa_i} \right)_{,j} \delta \omega_{ji} + \left[ \left( \dfrac{\partial H}{\partial E_i} \right)_{,i} + \rho_e \right] \delta \phi \right\} dV$$

$$+ \int_S \left\{ \left[ \dfrac{1}{2}\left( \dfrac{\partial H}{\partial e_{ij}} + \dfrac{\partial H}{\partial e_{ji}} \right) n_j - t_i^{(n)} \right] \delta u_i - \left[ \dfrac{1}{2}\dfrac{\partial H}{\partial \kappa_k} \varepsilon_{ijk} n_j + m_i^{(n)} \right] \delta \omega_i - \left[ \dfrac{\partial H}{\partial E_i} n_i + d \right] \delta \phi \right\} dS \quad \text{(A.11)}$$

At this stage, we notice that the rotation $\omega_i$ and displacement $u_i$ are related by

$$\omega_i = \dfrac{1}{2} \varepsilon_{ijk} u_{k,j} \quad \text{(A.12)}$$

As a result, we can continue with either of following alternative methods:

1. We directly use Eq. (A.12) in the variation Eq. (A.11), which results in the two fundamental governing Eqs. (24) and (25).
2. We use the Lagrange multiplier method to enforce the constraint Eq. (A.12), which results in the three fundamental governing Eqs. (16)-(18). This second approach follows the variational method developed in Darrall et al. (2014) for the purely mechanical consistent couple stress theory.

We demonstrate the details of these methods as follows.

## A.1. Direct method

We enforce the constraint (A.12) directly in the variation Eq. (A.11). As a result, we obtain

$$\delta \Pi_H = \int_V \left\{ \left[ -\frac{1}{2}\left(\frac{\partial H}{\partial e_{ij}} + \frac{\partial H}{\partial e_{ji}}\right)_{,j} + F_i \right] \delta u_i + \frac{1}{4}\left(\frac{\partial H}{\partial \kappa_i}\right)_{,j}(\delta u_{i,j} - \delta u_{j,i}) + \left[\left(\frac{\partial H}{\partial E_i}\right)_{,i} + \rho_e \right] \delta \phi_i \right\} dV$$
$$+ \int_S \left\{ \left[\frac{1}{2}\left(\frac{\partial H}{\partial e_{ij}} + \frac{\partial H}{\partial e_{ji}}\right) n_j - t_i^{(n)} \right] \delta u_i - \left[\frac{1}{2}\frac{\partial H}{\partial \kappa_k}\varepsilon_{ijk} n_j + m_i^{(n)} \right] \delta \omega_i - \left[\frac{\partial H}{\partial E_i} n_i + d \right] \delta \phi \right\} dS \quad (A.13)$$

or

$$\delta \Pi_H = \int_V \left\{ \left[ -\frac{1}{2}\left(\frac{\partial H}{\partial e_{ij}} + \frac{\partial H}{\partial e_{ji}}\right)_{,j} + F_i \right] \delta u_i + \frac{1}{4}\left[\left(\frac{\partial H}{\partial \kappa_i}\right)_{,j} - \left(\frac{\partial H}{\partial \kappa_j}\right)_{,i} \right] \delta u_{i,j} + \left[\left(\frac{\partial H}{\partial E_i}\right)_{,i} + \rho_e \right] \delta \phi_i \right\} dV$$
$$+ \int_S \left\{ \left[\frac{1}{2}\left(\frac{\partial H}{\partial e_{ij}} + \frac{\partial H}{\partial e_{ji}}\right) n_j - t_i^{(n)} \right] \delta u_i - \left[\frac{1}{2}\frac{\partial H}{\partial \kappa_k}\varepsilon_{ijk} n_j + m_i^{(n)} \right] \delta \omega_i - \left[\frac{\partial H}{\partial E_i} n_i + d \right] \delta \phi \right\} dS \quad (A.14)$$

This can be written as

$$\delta \Pi_H = \int_V \left\{ \begin{array}{l} \left[-\frac{1}{2}\left(\frac{\partial H}{\partial e_{ij}} + \frac{\partial H}{\partial e_{ji}}\right)_{,j} + F_i \right] \delta u_i + \left(\frac{1}{4}\left[\left(\frac{\partial H}{\partial \kappa_i}\right)_{,j} - \left(\frac{\partial H}{\partial \kappa_j}\right)_{,i} \right] \delta u_i \right)_{,j} \\ -\frac{1}{4}\left[\left(\frac{\partial H}{\partial \kappa_i}\right)_{,j} - \left(\frac{\partial H}{\partial \kappa_j}\right)_{,i} \right]_{,j} \delta u_i + \left[\left(\frac{\partial H}{\partial E_i}\right)_{,i} + \rho_e \right] \delta \phi_i \end{array} \right\} dV$$
$$+ \int_S \left\{ \left[\frac{1}{2}\left(\frac{\partial H}{\partial e_{ij}} + \frac{\partial H}{\partial e_{ji}}\right) n_j - t_i^{(n)} \right] \delta u_i - \left[\frac{1}{2}\frac{\partial H}{\partial \kappa_k}\varepsilon_{ijk} n_j + m_i^{(n)} \right] \delta \omega_i - \left[\frac{\partial H}{\partial E_i} n_i + d \right] \delta \phi \right\} dS \quad (A.15)$$

Once again we apply the divergence theorem to the second terms in the volume integral and recall the conditions $\delta u_i = 0$ on $S_u$, $\delta \omega_i = 0$ on $S_\omega$ and $\delta \phi = 0$ on $S_\phi$ to obtain the relation

$$\delta \Pi_H = \int_V \left\{ \begin{array}{l} \left[ -\frac{1}{2}\left(\frac{\partial H}{\partial e_{ij}} + \frac{\partial H}{\partial e_{ji}}\right)_{,j} + \frac{1}{4}\left[\left(\frac{\partial H}{\partial \kappa_i}\right)_{,j} - \left(\frac{\partial H}{\partial \kappa_j}\right)_{,i}\right]_{,j} + F_i \right] \delta u_i \\ + \left[\left(\frac{\partial H}{\partial E_i}\right)_{,i} + \rho_e \right] \delta \phi_i \end{array} \right\} dV$$

$$+ \int_{S_t} \left\{ \left[ \frac{1}{2}\left(\frac{\partial H}{\partial e_{ij}} + \frac{\partial H}{\partial e_{ji}}\right) + \frac{1}{4}\left[\left(\frac{\partial H}{\partial \kappa_i}\right)_{,j} - \left(\frac{\partial H}{\partial \kappa_j}\right)_{,i}\right] \right] n_j - \overline{t}_i^{(n)} \right\} \delta u_i dS \qquad (A.16)$$

$$- \int_{S_m} \left[ \frac{1}{2}\varepsilon_{ijk}\frac{\partial H}{\partial \kappa_k} n_j + \overline{m}_i^{(n)} \right] \delta \omega_i dS - \int_{S_d} \left[ \frac{\partial H}{\partial E_i} n_i + \overline{d} \right] \delta \phi dS = 0$$

In Eq. (A.16) the variations $\delta u_i$ and $\delta \phi$ are independent and arbitrary in the domain $V$. The variations of $\delta u_i$, $\delta \omega_i$ and $\delta \phi$ are also arbitrary on the boundary surfaces $S_t$, $S_m$ and $S_d$, respectively. Therefore, the individual terms in the integrals vanish separately and we have

$$\left[ \frac{1}{2}\left(\frac{\partial H}{\partial e_{ij}} + \frac{\partial H}{\partial e_{ji}}\right) + \frac{1}{4}\left[\left(\frac{\partial H}{\partial \kappa_i}\right)_{,j} - \left(\frac{\partial H}{\partial \kappa_j}\right)_{,i}\right] \right]_{,j} + F_i = 0 \quad \text{in } V \qquad (A.17)$$

$$-\left(\frac{\partial H}{\partial E_i}\right)_{,i} = \rho_e \quad \text{in } V \qquad (A.18)$$

$$\left[ \frac{1}{2}\left(\frac{\partial H}{\partial e_{ij}} + \frac{\partial H}{\partial e_{ji}}\right) + \frac{1}{4}\left[\left(\frac{\partial H}{\partial \kappa_i}\right)_{,j} - \left(\frac{\partial H}{\partial \kappa_j}\right)_{,i}\right] \right] n_j = \overline{t}_i^{(n)} \quad \text{on } S_t \qquad (A.19)$$

$$-\frac{1}{2}\varepsilon_{ijk}\frac{\partial H}{\partial \kappa_k} n_j = \overline{m}_i^{(n)} \quad \text{on } S_m \qquad (A.20)$$

$$-\frac{\partial H}{\partial E_i} n_i = \overline{d} \quad \text{on } S_d \qquad (A.21)$$

By comparing the boundary conditions Eqs. (A.19)-(A.21) with Eqs. (26)-(28), we obtain the general constitutive relations

$$\sigma_{(ji)} = \frac{1}{2}\left(\frac{\partial H}{\partial e_{ij}} + \frac{\partial H}{\partial e_{ji}}\right) \tag{A.22}$$

$$\mu_i = -\frac{1}{2}\frac{\partial H}{\partial \kappa_i} \tag{A.23}$$

$$D_i = -\frac{\partial H}{\partial E_i} \tag{A.24}$$

$$\sigma_{[ji]} = -\mu_{[i,j]} = \frac{1}{4}\left(\frac{\partial H}{\partial \kappa_i}\right)_{,j} - \frac{1}{4}\left(\frac{\partial H}{\partial \kappa_j}\right)_{,i} \tag{A.25}$$

Consequently, Eqs. (A.17) and (A.18) reduce to

$$[\sigma_{(ji)} + \mu_{[j,i]}]_{,j} + F_i = 0 \tag{A.26}$$

$$D_{i,i} = \rho_e \tag{A.27}$$

Therefore, we can see that the variational formulation has produced the fundamental governing Eqs. (24) and (25).

### A.2. Lagrange multiplier method

We can enforce the constraint (A.12) by using the Lagrange multiplier method. Accordingly, we define a new functional

$$\tilde{\Pi}_H(u,e,\kappa,\lambda) = \Pi_H(u,e,\kappa) + \int_V \lambda_i\left(\omega_i - \frac{1}{2}\varepsilon_{ijk}u_{k,j}\right)dV \tag{A.28}$$

where the components of $\lambda_i$ are the Lagrange multiplier functions.

Therefore, for the first variation of the functional $\tilde{\Pi}_H$ we have

$$\delta\tilde{\Pi}_H = \delta\Pi_H + \int_V \left[\lambda_i\delta\omega_i - \frac{1}{2}\lambda_i\varepsilon_{ijk}\delta u_{k,j} + \delta\lambda_i\left(\omega_i - \frac{1}{2}\varepsilon_{ijk}u_{k,j}\right)\right]dV \tag{A.29}$$

By some manipulation, we obtain

$$\delta \tilde{\Pi}_H = \delta \Pi_H + \int_V \left[ \lambda_i \delta \omega_i - \frac{1}{2}\left(\varepsilon_{ijk}\lambda_i \delta u_k\right)_{,j} + \frac{1}{2}\varepsilon_{ijk}\lambda_{i,j}\delta u_k + \delta\lambda_i\left(\omega_i - \frac{1}{2}\varepsilon_{ijk}u_{k,j}\right) \right]dV \qquad (A.30)$$

Now by applying the divergence theorem to the second volume integral, we obtain

$$\delta \tilde{\Pi}_H = \delta \Pi_H + \int_V \left[ \lambda_k \delta \omega_k - \frac{1}{2}\varepsilon_{ijk}\lambda_{k,j}\delta u_i + \delta\lambda_i\left(\omega_i - \frac{1}{2}\varepsilon_{ijk}u_{k,j}\right) \right]dV$$
$$+ \int_S \frac{1}{2}\varepsilon_{ijk}\lambda_k \delta u_i n_j dS \qquad (A.31)$$

Then by using Eq. (A.11), this variation can be written as

$$\delta \tilde{\Pi}_H = \int_V \left\{ \begin{array}{l} \left[-\frac{1}{2}\left(\frac{\partial H}{\partial e_{ij}} + \frac{\partial H}{\partial e_{ji}}\right)_{,j} + \varepsilon_{ijk}\lambda_{k,j} + F_i\right]\delta u_i - \left[\frac{1}{2}\left(\frac{\partial H}{\partial \kappa_i}\right)_{,j}\varepsilon_{ijk} - \lambda_k\right]\delta\omega_k \\ + \left[\left(\frac{\partial H}{\partial E_i}\right)_{,i} + \rho_e\right]\delta\phi_i \end{array} \right\}dV$$
$$+ \int_S \left\{\left[\frac{1}{2}\left(\frac{\partial H}{\partial e_{ij}} + \frac{\partial H}{\partial e_{ji}}\right) + \varepsilon_{ijk}\lambda_k\right]n_j - t_i^{(n)}\right\}\delta u_i dS - \int_S \left[\frac{1}{2}\frac{\partial H}{\partial \kappa_k}\varepsilon_{ijk}n_j + m_i^{(n)}\right]\delta\omega_i dS \qquad (A.32)$$
$$- \int_S \left[\frac{\partial H}{\partial E_i}n_i + d\right]\delta\phi dS + \int_V \delta\lambda_i\left(\omega_i - \frac{1}{2}\varepsilon_{ijk}u_{k,j}\right)dV$$

Now by imposing the conditions $\delta u_i = 0$ on $S_u$, $\delta \omega_i = 0$ on $S_\omega$ and $\delta\phi = 0$ on $S_\phi$, we obtain the relation

$$\delta\tilde{\Pi}_H = \int_V \left\{ \begin{array}{l} \left[ -\left[ \dfrac{1}{2}\left( \dfrac{\partial H}{\partial e_{ij}} + \dfrac{\partial H}{\partial e_{ji}} \right) + \varepsilon_{ijk}\lambda_{k,j} + F_i \right] \delta u_i - \left[ \dfrac{1}{2}\left( \dfrac{\partial H}{\partial \kappa_i} \right)_{,j} \varepsilon_{ijk} - \lambda_k \right] \delta\omega_k \right. \\ \left. + \left[ \left( \dfrac{\partial H}{\partial E_i} \right)_{,i} + \rho_e \right] \delta\phi_i \end{array} \right\} dV$$

$$+ \int_{S_t} \left\{ \left[ \dfrac{1}{2}\left( \dfrac{\partial H}{\partial e_{ij}} + \dfrac{\partial H}{\partial e_{ji}} \right) + \varepsilon_{ijk}\lambda_k \right] n_j - \overline{t}_i^{(n)} \right\} \delta u_i dS - \int_{S_m} \left[ \dfrac{1}{2}\dfrac{\partial H}{\partial \kappa_k} \varepsilon_{ijk} n_j + \overline{m}_i^{(n)} \right] \delta\omega_i dS \quad \text{(A.33)}$$

$$- \int_{S_d} \left[ \dfrac{\partial H}{\partial E_i} n_i + \overline{d} \right] \delta\phi dS + \int_V \delta\lambda_i \left( \omega_i - \dfrac{1}{2}\varepsilon_{ijk} u_{k,j} \right) dV$$

In Eq. (A.33) the variations $\delta u_i$, $\delta\omega_i$ and $\delta\phi$ are independent and arbitrary in the domain $V$. These variations are also arbitrary on the boundary surfaces $S_t$, $S_m$ and $S_d$, respectively. Therefore, the individual terms in the integrals vanish separately and we have

$$\left[ \dfrac{1}{2}\left( \dfrac{\partial H}{\partial e_{ij}} + \dfrac{\partial H}{\partial e_{ji}} \right) + \varepsilon_{ijk}\lambda_k \right]_{,j} + F_i = 0 \quad \text{in } V \quad \text{(A.34)}$$

$$\dfrac{1}{2}\left( \dfrac{\partial H}{\partial \kappa_i} \right)_{,j} \varepsilon_{ijk} - \lambda_k = 0 \quad \text{in } V \quad \text{(A.35)}$$

$$-\left( \dfrac{\partial H}{\partial E_i} \right)_{,i} = \rho_e \quad \text{in } V \quad \text{(A.36)}$$

$$\left[ \dfrac{1}{2}\left( \dfrac{\partial H}{\partial e_{ij}} + \dfrac{\partial H}{\partial e_{ji}} \right) + \varepsilon_{ijk}\lambda_k \right] n_j = \overline{t}_i^{(n)} \quad \text{on } S_t \quad \text{(A.37)}$$

$$-\dfrac{1}{2}\varepsilon_{ijk}\dfrac{\partial H}{\partial \kappa_k} n_j = \overline{m}_i^{(n)} \quad \text{on } S_m \quad \text{(A.38)}$$

$$-\dfrac{\partial H}{\partial E_i} n_i = \overline{d} \quad \text{on } S_d \quad \text{(A.39)}$$

$$\omega_i = \dfrac{1}{2}\varepsilon_{ijk} u_{k,j} \quad \text{in } V \quad \text{(A.40)}$$

By comparing the boundary conditions Eqs. (A.37)-(A.39) with Eqs. (26)-(28), we identify the Lagrange multiplier vector function $\lambda_i$ as the vector $s_i$, which represents the skew-symmetric part of the force stress tensor $\sigma_{[ji]}$ (Darrall et al., 2014), that is

$$\lambda_k = s_k = \frac{1}{2}\varepsilon_{ijk}\sigma_{[kj]} \tag{A.41}$$

$$\sigma_{[ji]} = \varepsilon_{ijk}\lambda_k \tag{A.42}$$

As a result, we obtain the general constitutive relations

$$\sigma_{(ji)} = \frac{1}{2}\left(\frac{\partial H}{\partial e_{ij}} + \frac{\partial H}{\partial e_{ji}}\right) \tag{A.43}$$

$$\mu_i = -\frac{1}{2}\frac{\partial H}{\partial \kappa_i} \tag{A.44}$$

$$D_i = -\frac{\partial H}{\partial E_i} \tag{A.45}$$

$$\sigma_{[ji]} = -\mu_{[i,j]} = \frac{1}{4}\left(\frac{\partial H}{\partial \kappa_i}\right)_{,j} - \frac{1}{4}\left(\frac{\partial H}{\partial \kappa_j}\right)_{,i} \tag{A.46}$$

Consequently, Eqs. (A34)-(A36) reduce to

$$\sigma_{ji,j} + F_i = 0 \qquad \text{Linear balance law} \tag{A.47}$$

$$\mu_{ji,j} + \varepsilon_{ijk}\sigma_{jk} = 0 \qquad \text{Angular balance law} \tag{A.48}$$

$$D_{i,i} = \rho_e \qquad \text{Gauss law} \tag{A.49}$$

It is seen that this variational formulation has produced the three fundamental governing equations (16)-(18).